\documentclass[conference]{sig-alternate-05-2015}

\usepackage{alltt                                    % I like these
          , multirow
          , booktabs
          , listings
          , graphicx
          ,float
	,cite
          ,verbatim
         ,mathtools
	,url
	,amsmath
}
\usepackage[table]{xcolor}
\usepackage[numbers]{natbib}     % this is a better citation system
\usepackage{syntax}
\usepackage{algorithm}
\usepackage{enumitem}
\usepackage{threeparttable}
\usepackage{framed}
\usepackage{algpseudocode}
\usepackage{hhline}
\usepackage{balance}
\usepackage{setspace}
\usepackage{afterpage}

\usepackage{expl3}
\ExplSyntaxOn
\newcommand\latinabbrev[1]{
  \peek_meaning:NTF . {% Same as \@ifnextchar
    #1\@}%
  { \peek_catcode:NTF a {% Check whether next char has same catcode as \'a, i.e., is a letter
      #1., \@ }%
    {#1., \@}}}
\ExplSyntaxOff

\newcommand{\CASE}[1]{\STATE \textbf{case} #1\textbf{:} \begin{ALC@g}}
\newcommand{\ENDCASE}{\end{ALC@g}}

\newcommand{\DEFAULT}{\STATE \textbf{default:} \begin{ALC@g}}
\newcommand{\ENDDEFAULT}{\end{ALC@g}}
\newcommand{\DEFAULTLINE}[1]{\STATE \textbf{default:} }
\let\footnotesize\scriptsize

\algnewcommand{\LineComment}[1]{\State \(\triangleright\) #1}

\newlength{\oldtextfloatsep}

\newsavebox{\supbox}% Superscript box
\newcommand{\bsup}{\begin{lrbox}{\supbox}$\tt\scriptstyle}% Superscript begin
\newcommand{\esup}{$\end{lrbox}{}^{\usebox{\supbox}}}% Superscript end
\def\eg{\latinabbrev{e.g}}
\def\ie{\latinabbrev{i.e}}

\definecolor{lightpurple}{rgb}{0.8,0.8,1}
\definecolor{codebg}{RGB}{255,255,255}
\definecolor{commentcolor}{RGB}{11,140,11}
\begin{document}

%\setcopyright{acmcopyright}
%\isbn{978-1-4503-4205-6}
%\conferenceinfo{ICSE}{'2016 Austin, Texas USA}

\CopyrightYear{2016} 
\setcopyright{acmcopyright}
\conferenceinfo{ICSE '16 Companion,}{May 14-22, 2016, Austin, TX,
USA}
\isbn{978-1-4503-4205-6/16/05}\acmPrice{\$15.00}
\doi{http://dx.doi.org/10.1145/2889160.2889244}

\title{CORRECT: Code Reviewer Recommendation in GitHub Based on Cross-Project and Technology Experience}

%\author{\IEEEauthorblockN{Mohammad Masudur Rahman  ~~~ Chanchal K. Roy ~~~ $^\dagger$Jason Collins}
%\IEEEauthorblockA{University of Saskatchewan, Canada,  $^\dagger$Google Inc., USA\\
%\{masud.rahman, chanchal.roy\}@usask.ca,  $^\dagger$jasonco@google.com}
%}

\numberofauthors{1} %  in this sample file, there are a *total*
\author{
% 1st. author
\alignauthor
Mohammad Masudur Rahman~~~Chanchal K. Roy~~~$^\dagger$Jason A. Collins\\
       \affaddr{University of Saskatchewan, Canada, $^\dagger$Google Inc., USA}\\
      \email{\{masud.rahman, chanchal.roy\}@usask.ca, $^\dagger$jasonco@google.com}
% 2nd. author
%\alignauthor
%Chanchal K. Roy\\
%       \affaddr{University of Saskatchewan, Canada}\\
%\affaddr{chanchal.roy@usask.ca}
%% 3rd. author
%\alignauthor
%$^\dagger$Jason Collins\\
%       \affaddr{$^\dagger$Google Inc., USA}\\
%\affaddr{$^\dagger$jasonco@google.com}
% 3rd. author
}

\maketitle

\begin{abstract}
%Code review is considered as one the most important developer collaborations in modern collaborative software development. 
Peer code review locates common coding rule violations and simple logical errors in the early phases of software development, and thus reduces overall cost. 
However, in GitHub, identifying an appropriate code reviewer for a pull request is a non-trivial task given that reliable information for reviewer identification is often not readily available. 
In this paper, we propose a code reviewer recommendation technique that considers not only the relevant cross-project work history (\eg\ external library experience) but also 
the experience of a developer in certain specialized technologies associated with a pull request
for determining her expertise as a potential code reviewer.
We first motivate our technique using an exploratory study with 10 commercial projects and 10 associated libraries external to those projects.
Experiments using 17,115 pull requests from 10 commercial projects and six open source projects show that our technique provides 85\%-- 92\% recommendation accuracy, about 86\% precision and 79\%--81\% recall in code reviewer recommendation, which are highly promising.
Comparison with the state-of-the-art technique also validates the empirical findings and the superiority of our recommendation technique.
%Information on developer's expertise is not often readily available, and reliable estimation of such expertise is trivial. 
%Our technique is found highly promising not only for closed-source projects but also for open source projects, and it 
%The technique also outperforms the state-of-the-art techniques.
%its effectiveness in recommending appropriate developers for code reviews.
\end{abstract}

%A category including the fourth, optional field follows...
%\category{D.2.8}{Software Engineering}{Heuristics}[cross-project experience, specialized technology experience]




%\begin{keywords}
%Code reviewer recommendation, cross-project experience, technology experience, GitHub, pull request  
%\end{keywords}

\section{Introduction}
Software development practices have dramatically changed over the last decade, and software projects are now developed not only in a collaborative environment but also in a distributed fashion \cite{seafood}.
GitHub, a collaborative and distributed development framework, promotes pull-request based development where a new developer forks from an existing repository (\ie\ project), works on certain module of her interest, and then submits the changed files to the repository using a pull request \cite{pull}. 
%The request is then carefully analyzed, and 
One or more expert developers from the base repository, referred by the submitter, then review(s) the code carefully before accepting the changes as a contribution to the codebase.
%The administrators of that repository then review the changes, consults with the developer, and accepts the contribution if satisfied.
Such code review is reported as highly effective for locating common coding rule violations or for performing simple logical verifications \cite{vcc,expect,contempo}.
It also helps identify the issues (\eg\ vulnerabilities) in the code in the early phases of development, and thus reduces overall cost for the software project \cite{vcc,expect}. 
However, choosing an appropriate developer for code review for a pull request is a significant challenge \cite{reduce}, and to date, GitHub does not provide any support for this.
Reliable information on developer's expertise (\eg\ technology skill) for the review is often not readily available, and it needs to be carefully mined from the codebase.
Thus, reviewer identification task is even more challenging and time-consuming for the novice developers who are less familiar with the codebase as well as the skills of the hundreds of fellow developers.
%They often experience difficulties in identifying the code reviewers from hundreds of project members.
Such challenge is prevalent not only in open source development but also in
the industrial environment where a company uses GitHub for commercial development, and encourages developer collaborations such as peer code review.
%New developers working on a project 
%Fortunately, an automated support in identifying the appropriate developers for code review can greatly help in this regard.

Fortunately, there have been several studies that recommend code reviewers by analyzing past code review history (\eg\ line change history \cite{reduce}, review comments \cite{yu,xin}), project directory structure \cite{pick, rveffect}, and developer collaboration network \cite{yu}.  
Similarly, studies on expert recommendation for software bugs also exploit different software artifacts \cite{kevic,authorship} and developer communication history \cite{sna}.
%version control history \cite{kevic, bosu}, authors' meta data \cite{authorship, discovery}, and developer communication history \cite{brazil, seafood, sna, ghadeer}, and perform static analysis such as code similarity analysis \cite{ghadeer}.
%In short, 
Thus, existing studies mostly rely on the work history of a developer within a particular project and her collaboration history with other developers for determining her expertise.  
However, no studies consider the cross-project experience or the experience in various specialized technologies of a developer, and thus they fall short in handling certain challenges.

First, in the industry, software developers often reuse software components (\eg\ libraries) that are previously developed by themselves for low cost and faster development.
Thus, their contributions scatter throughout different projects in the code repositories of the organization, and such contributions are a great proxy to their experience. 
Unfortunately, the existing studies on code reviewer recommendation completely ignore such information in expertise determination, and their recommendations are merely based on the contribution details within a particular project.
Second, underlying tools and technologies of software projects are rapidly changing, and modern projects often involve different  specialized and cutting edge technologies such as \texttt{map-reduce, task queues, urlfetch, memcache} and \texttt{pipeline}.
Hence, code reviewers for a pull request are expected to have expertise in such technologies.
However, neither mining of the revision history of changed files nor mining of the developer collaboration history, as the existing studies do, might be sufficient enough to ensure that.
%One possible way is to analyze the past experience of a developer with those technologies.
%an analysis on the cross-project work experience of the developers is essential to determine thier relevant expertise for a pull request.
%Third, code reviewer recommendation entirely based on version history analysis might not be effective in an organizational context if the recommended developers are unavailable (\ie\ already left the organization) or overloaded with review requests.
%Third, recently proposed tools are reported as effective in identifying simple coding rule violations \cite{reduce, panichella}, and thus, human reviewers are expected to identify potential logical concerns such as security vulnerabilities \cite{vcc} or performance bottlenecks \cite{expect}.
%Such identifications often require special skills or application domain expertise, and previous developers of the changed files might always not be the most appropriate candidates for a code review. 
Thus, a technique that can analyze both relevant cross-project experience and specialized technology experience of a developer for a pull request, is likely to overcome the above challenges.
% associated with the recommendation of code reviewers. 

%and identifies one or more experts from a list of available developers, is likely to solve our research problem.
%\begin{figure*}[!t]
%\centering
%\includegraphics[width=6in]{sysdiag5}
%\vspace{-.4cm}
%\caption{Schematic Diagram of Proposed Approach}
%\vspace{-.4cm}
%\label{fig:sysdiag}
%\end{figure*}

In this paper, we propose a novel code reviewer recommendation technique--CoRReCT 
(\textbf{Co}de \textbf{R}eviewer \textbf{Re}commen-dation based on \textbf{C}ross-project and \textbf{T}echnology experience),
 for pull requests at GitHub. 
It estimates code review expertise of a developer for a pull request by analyzing her past work experience with (1) external software libraries and (2) specialized technologies used by the pull request.
Reference to the external libraries (\ie\ software units external to the working project) in the code generally suggests one's working experience with such libraries, and we call it \emph{cross-project experience}.
Our baseline idea is-- \emph{``if a past pull request uses similar external libraries or similar specialized technologies to the current pull request, then the past request is relevant to the current request, and thus, its reviewers are also potential candidates for the code review of the current request"}.
We first mine the library and technology information from a pull request using static analysis, and then identify the relevant requests in terms of library and technology similarities from the recently submitted request collection.
%Since such similarity measures estimate the relevance between pull requests, we propagate those measures to corresponding reviewers for heuristically capturing their expertise for the current pull re
We then propagate the similarity score for each relevant request to its corresponding code reviewers as a proxy to the shared experience in external libraries and specialized technologies with the current request.
Thus, each of the candidates accumulates scores for all relevant requests, and finally, the technique returns a ranked list of code reviewers.
While the technique adopts heuristics for ranking, our contribution lies in identifying the appropriate proxies to code review expertise.
%We then exploit the relevance estimates between the current and the past pull requests to heuristically estimate 
%and propagate the similarity estimates to their corresponding code reviewers.
 %by analyzing not only her familiarity with different
%technologies associated with a pull request but also her past experience with different software libraries (previously developed) included in the source files of the request.  
%cross-project experience within the codebase. 
%It then ranks the candidate developers based on two experience dimensions-- \emph{cross-project experience} and \emph{technology experience}, and recommends the top-ranked developers as code reviewers for the request.
%Thus, the baseline idea is that the past developers who have more experience with the  libraries included or technologies adopted in the changed files of a given pull request, are more appropriate for code review than the ones with less experience.
%To the best of our knowledge, no studies so far consider either cross-project experience or specialized technology experience in code reviewer recommendation.
To the best of our knowledge, ours is the first study that exploits the benefits of using cross-project experience and specialized technology experience (of the developers) in recommending code reviewers for a pull request.
%technology experience or cross-project experience as an expertise dimension despite of their potential for code reviewer recommendation.
We adopt a client-server model for our technique where the client module is packaged as a Google Chrome plug-in, and the server module is hosted as a web service. 
Both modules are available online \cite{correct} for replication or third party use.
%\footnote{http://www.usask.ca/$\sim$masud.rahman/correct}.

In order to motivate cross-project experience and specialized technology experience as a proxy to code review expertise, we first conducted an exploratory study using 10 commercial projects and 10 libraries from the codebase of 
a local reputed software company (for the sake of anonymity, we call this \emph{ABC}) with more than 150 employees. 
%\emph{VendAsta Technologies}, a local reputed software company with more than 150 employees.
Experiments with 10 commercial projects and six open source projects totaling 17,115 pull requests show that our technique recommends code reviewers with 85\%-- 92\% recommendation accuracy, about 86\% precision and 79\%--81\% recall.
Furthermore, comparison with the state-of-the-art also confirms the empirical findings and superiority of our technique.
%Experiments using 17,115 pull requests from 10 commercial projects and six open source projects show that our technique provides 85\%-- 92\% recommendation accuracy, about 86\% precision and 79\%--81\% recall in code reviewer recommendation, which are highly promising.
%Our technique is found highly promising not only for closed-source projects but also for open source projects, and it 
%Comparison with the state-of-the-art technique and a user study involving 16 professional developers also confirm our empirical findings and the good usability of the technique.
%Experiments using 13,081 pull requests from the commercial projects and comparison with a state-of-the-art technique show that our proposed technique is highly promising.  
%The technique recommends code reviewers with a recommendation accuracy of 92.15\%, a precision of 85.93\% and a recall of 81.39\%, which considerably outperforms the state-of-the-art techniques from the literature.
%We also validate our findings using 4,034 pull requests from six open source projects, and a user study involving 16 professional developers.
%carefully apply different statistical tests, and discuss the implications of our findings.
Thus, we make the following contributions in the paper:
\begin{itemize}[noitemsep,topsep=1pt]
\item An exploratory study that not only analyzes the usage of external libraries and specialized technologies in the commercial projects but also investigates their potential for code review.
\item Two novel expertise dimensions--\emph{cross-project experience} and \emph{specialized technology experience} for code reviewer recommendation for pull requests at GitHub.
\item Comprehensive evaluation of the proposed technique with both commercial and open source projects using popular performance metrics, and comparison with the state-of-the-art.
\item Implementation of our recommendation technique as a web service (server) and a Chrome plug-in (client). 
\end{itemize}

%First, we analyze code review and version control history of each of the existing repositories in the codebase, and extract review and authorship contributions of the developers. 
%Second, we analyze each repository down to class or method level, identify different technological aspects (\eg\ \emph{database transactions, map-reduce, urlfetch}) through a set of NLP algorithms adapted for source code, and then develop a \emph{global index} for such technological concepts against the source files. 
%The idea is to capture technology domain-specific and cross-project expertise of a developer using IR-based techniques.
%%Second, we analyze code review and version control history of the codebase, extract file and authorship level metrics, and then combine them with the expertise metrics from previous step in order to develop a \emph{heuristic expertise model}. 
%Third, we combine different metrics collected from previous steps, develop an expertise model, and 
%evaluate the model using a case study. We use the pull request and code review data of VendAsta Technologies for the case study.
%Fourth, we implement the recommendation system as a Google Chrome plug-in.

The rest of the paper is structured as follows-- Section \ref{sec:mcr} provides an overview on modern code reviw, Section \ref{sec:explo} focuses on our conducted exploratory study, and Section \ref{sec:proposed} describes our proposed recommendation technique.
Section \ref{sec:experiment} discusses the evaluation and validation details, and Section \ref{sec:threats} focuses on threats to the validity of our findings.  
Section \ref{sec:related} discusses related studies from the literature, and finally Section \ref{sec:conclusion} concludes the paper with future work.

% Thus the paper makes the following technical contributions:
%\begin{itemize}
%\item We propose a heuristic model that estimates expertise of a developer by analyzing source file level, technology specific and cross-repository contributions of the developer.
%\item We report a case study with XXX pull requests from \emph{VendAsta} repositories where we evaluate our proposed expertise model in recommending appropriate developers for code reviews.
%\end{itemize}

\begin{figure}[!t]
\centering
\includegraphics[width=3in]{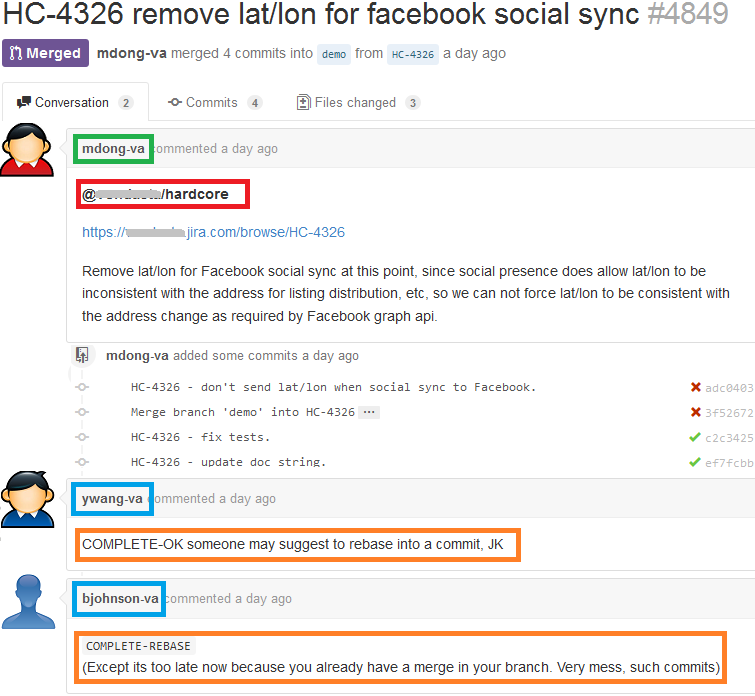}
\vspace{-.3cm}
\caption{Code review interface at GitHub}
\vspace{-.4cm}
\label{fig:mcr}
\end{figure}

%\begin{figure*}[!t]
%\centering
%\includegraphics[width=5in]{depen2}
%\vspace{-.3cm}
%\caption{Dependency of VendAsta projects on the selected libraries}
%\vspace{-.4cm}
%\label{fig:depen}
%\end{figure*}

%\section{Background}\label{sec:bg}
\section{Modern Code Review}\label{sec:mcr}
\emph{Code review} refers to a manual assessment of source code that identifies potential defects (\eg\ logical errors) and quality problems (\eg\ coding rule violations) in the code \cite{modern}. 
%There exist different ways for code review including the most formal \emph{code inspection} to the least formal \emph{over-the-shoulder} review \cite{reduce}.
In recent years, code review is assisted with different tools, which is less formal and more popular than the traditional review techniques \cite{reduce}. Such code review is termed as \emph{Modern Code Review} (MCR) \cite{modern}. 
It is widely adopted both by the commercial organizations (\eg\ Google, Microsoft) and by the open-source communities (\eg\ Android, LibreOffice).
Existing code review tools such as \emph{ReviewBot} \cite{reduce} and \emph{RevFinder} \cite{pick} are mostly based on Gerrit, a web-based code review system.
%\footnote{Web based code review and project management for Git based projects}.
GitHub also provides a similar feature for conducting code review by human developers through pull requests where a developer can request her peers for code review during a pull request submission.  

For example, developer \emph{mdong-va} (\ie\ green box, Fig. \ref{fig:mcr}) requests a development team-- \emph{hardcore} (\ie\ red box, Fig. \ref{fig:mcr}) for code review during the submission of pull request \#4849.
Two developers--\emph{ywang-va} and \emph{bjohnson-va} (\ie\ blue boxes) from the team analyze the commits associated with the pull request, perform the code review, and then post their feedback using comments (\ie\ orange boxes, Fig. \ref{fig:mcr}). 
%The submitter then addresses the comments, and resubmits the changed code. 
%Once the reviewers are satisfied and add the \emph{COMPLETE-OK} label to the reviewed code (\ie\ orange boxes, Fig. \ref{fig:mcr}), the pull request is merged into the codebase.
Unfortunately, despite assistance from the static analysis tools \cite{reduce}, effective code review still remains a challenge,
and identifying an appropriate reviewer is a non-trivial task.
To date, both reviewer selection and code review are also performed manually at GitHub.
In this research, we thus recommend appropriate developers (\eg\ \emph{ywang-va, bjohnson-va}) for such code review task (\eg\ Fig. \ref{fig:mcr}) at GitHub.

\section{Exploratory Study}\label{sec:explo}
%In the industry, software developers often reuse previously developed software components (\eg\ libraries) and adopt various specialized technologies (\eg\ \texttt{mapreduce, pipleline}) for low cost and faster development.
%However, in order to better understand their usage patterns in the projects, and to investigate the potential of developer experience with such libraries and technologies for code review, an exploratory study is warranted. 
%In this section, we conduct an exploratory study using 20 commercial projects and libraries and considering ten specialized technologies. In our study, we answer three research questions as follows: 
Since our proposed technique is based on cross-project work history (\ie\ external library experience) and specialized technology experience, it is important that we first conduct an exploratory study to find out to what extent such situations in fact occur and whether such information can help recommend code reviewers. 
We thus conduct an exploratory study with 20 commercial projects and libraries targeting 10 specialized technologies.
In our study, we answer three research questions as follows: 

\begin{itemize}[noitemsep,topsep=1pt]
\item \textbf{Exp-RQ$\mathbf{_1}$:} How frequently do the commercial software projects reuse libraries from the codebase?
\item \textbf{Exp-RO$\mathbf{_2}$:} Does the experience of a developer with such libraries matter in code reviewer selection?
\item \textbf{Exp-RQ$\mathbf{_3}$:} How frequently do the commercial software projects use specialized technologies?
\end{itemize}

\subsection{Dataset Collection}\label{sec:dscoll}
ABC codebase hosts 30 commercial projects and 21 libraries (according to March, 2015) having various sizes and functionalities. They are based on \emph{Google Cloud} platform and are mostly written in \emph{Python}.
In order to perform a meaningful analysis on the codebase, we select (1) the top 10 projects with more than 750 closed pull requests and (2) the top 10 libraries that are used at least 10 times on average in each of those projects for the exploratory study. 
Table \ref{table:project} and Table \ref{table:library} show the details of the selected projects and libraries respectively.
We also consider 10 specialized technologies that are at least used five times on average in each of those projects for the study. 
Table \ref{table:technology} shows the selected technologies, and their specialized functionalities.

\begin{table}
\centering
\caption{ABC Company Projects}\label{table:project}
%\vspace{-.2cm}
\resizebox{3.3in}{!}{%
\begin{threeparttable}
\begin{tabular}{l|c|c|c||l|c|c|c}
\hline
\textbf{Project} & \textbf{\#Files}\tnote{1} & \textbf{\#PR}\tnote{2} & \textbf{\#PRR}\tnote{3}& \textbf{Project} & \textbf{\#Files} & \textbf{\#PR} & \textbf{\#PRR}\\
\hline
CS & 3,733 & 4,560 & 57 & SR & 2,139 & 1,927 & 36\\
\hline
ARM & 2,035 & 969 & 33 & NB &1,524 & 828 & 32 \\
\hline
SM & 2,026 & 1,291 & 36 & VBC & 1,894& 1,050 & 36 \\
\hline
VW & 1,475 & 787 & 16 & AA & 2,174 & 1,313 & 46 \\
\hline
MS & 2,227 & 1,156 & 36 & ST & 2,676 & 1,397 & 28\\
\hline
\end{tabular}
%\begin{tablenotes}
%\item [1] No. of example pairs for which relative quality evaluation matches with that of StackOverflow
%\item [2] \% of agreement, \item [3] \% of disagreement
 %\end{tablenotes}
\center
$^1$Source files, $^2$Pull requests, $^3$Pull request reviewers
\end{threeparttable}
%\vspace{-.2cm}
}
\vspace{-.6cm}
\end{table}

\begin{table}
\centering
\caption{ABC Company Libraries}\label{table:library}
%\vspace{-.2cm}
\resizebox{3.3in}{!}{%
\begin{threeparttable}
\begin{tabular}{l|c|c|c||l|c|c|c}
\hline
\textbf{Library} & \textbf{\#Files}\tnote{1} & \textbf{\#TC}\tnote{2} & \textbf{\#TA}\tnote{3}& \textbf{Library} & \textbf{\#Files} &  \textbf{\#TC} & \textbf{\#TA}\\
\hline
vapi & 631 & 727 & 24 & vpubsub &750 &428 &20\\
\hline 
 vform & 545 &893 &27 &vtest & 511&222&18\\
\hline
vbackup & 469 & 236 & 15 &vauth &421 &200 &19\\
\hline
vlogs & 826&243 & 17 &vmonitor & 532&213 &12\\
\hline
vautil & 1294& 269 & 20 &vpipeline & 1470 &228 &13\\
\hline
\end{tabular}
\center
$^1$Source files, $^2$Total commits, $^3$Total authors
%\begin{tablenotes}
%\item [1] No. of example pairs for which relative quality evaluation matches with that of StackOverflow
%\item [2] \% of agreement, \item [3] \% of disagreement
 %\end{tablenotes}
\end{threeparttable}
%\vspace{-.2cm}
}
\vspace{-.6cm}
\end{table}

\begin{table}[!t]
\centering
\caption{Specialized Technologies in ABC Projects}\label{table:technology}
%\vspace{-.2cm}
\resizebox{3.3in}{!}{%
\begin{threeparttable}
\begin{tabular}{l|l||l|l}
\hline
\textbf{Technology} & \textbf{Functionality} & \textbf{Technology} & \textbf{Functionality}\\
\hline
\texttt{taskqueue} & task scheduling & \texttt{deferred} & task scheduling \\
\hline
\texttt{mapreduce} & distributed computing & \texttt{blobstore} & data storage\\
\hline
\texttt{urlfetch} & HTTP communication & \texttt{jinja2} & template engine \\
\hline
\texttt{search} & item search& \texttt{modules} & app. factorization\\
\hline
\texttt{ndb} &data storage & \texttt{socket} & networking\\
\hline
\end{tabular}
%\begin{tablenotes}
%\item [1] No. of example pairs for which relative quality evaluation matches with that of StackOverflow
%\item [2] \% of agreement, \item [3] \% of disagreement
 %\end{tablenotes}
\end{threeparttable}
%\vspace{-.2cm}
}
\vspace{-.4cm}
\end{table}

\subsection{Answering Exp-RQ$\mathbf{_1}$: Frequency of library use in the commercial projects}
%From Fig. \ref{fig:depen}, we note that our selected projects are highly dependent on various commercial libraries from the code repository.
%In fact, 
Each of the selected projects nearly depends on all of the chosen libraries for their functionalities (\eg\ authentication, utility).
However, in order to answer Exp-RQ$_1$, we need more focused statistics which are provided using the following analyses both with source files and pull requests. 

\textbf{Library Use in Source Files:} In Python programming, external libraries are generally attached to a source file using import statements.
We analyze the import statements (\eg\ \texttt{import vform, from vlogs import AbstractTracer}) from all the source files of each of the selected projects, and extract the library names (\eg\ \texttt{vform, vlogs}) using a custom-built Python AST parser.
The goal is to determine the extent to which a particular external library (\eg\ \texttt{vform}) is used in different projects. 

\begin{figure}[!t]
\centering
\includegraphics[width=3.2in]{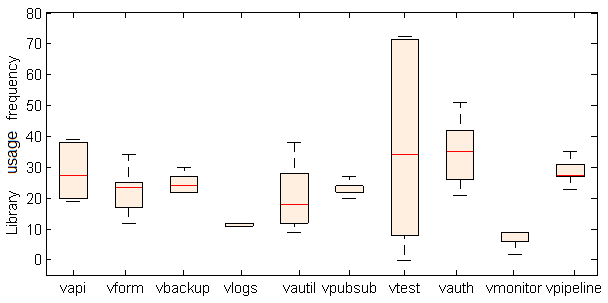}
\vspace{-.5cm}
\caption{Usage frequency of libraries (Table \ref{table:library}) in ABC Company projects (Table \ref{table:project})}
\vspace{-.3cm}
\label{fig:libfreq}
\end{figure}

Fig. \ref{fig:libfreq} shows the box plot of frequency of occurrence in different commercial projects for the selected libraries.
We note that \emph{vtest}, a testing library, has a large variance in usage frequency with a median frequency around 35 and a maximum frequency around 70.
This means that the library is used to various degrees in different projects for testing purposes. 
The similar median also goes for \emph{vauth} library (\ie\ provides authentication support) whereas 
the other libraries have a median frequency around 25. 
Thus, each project is densely connected with most of the libraries, the overall dependency of the projects on the external libraries is remarkable due to their dedicated functionalities.

%We also investigate the relative usage frequency for the selected libraries in different projects. Fig. \ref{fig:libstats} shows the heat map of frequency that demonstrates 
%the comparative use of each library in different software projects.
%We note that \emph{vtest} and \emph{vauth} are extensively referred to by most of the projects which confirms our empirical findings in Fig. \ref{fig:libfreq}.
%The other libraries--\emph{vapi, vform, vbackup, vautil, vpubsub} and \emph{vpipeline} are also used moderately.
%Thus, most of the selected libraries (Table \ref{table:library}) are used considerably in the software projects under study, which suggests that the usage of external libraries with dedicated functionalities is a major aspect of commercial software development.

%\begin{figure}[!t]
%\centering
%\includegraphics[width=3.3in]{libstats2}
%\vspace{-.7cm}
%\caption{Relative frequency of the use of libraries (Table \ref{table:library}) in VendAsta projects (Table \ref{table:project})}
%\vspace{-.3cm}
%\label{fig:libstats}
%\end{figure}

\begin{figure}[!t]
\centering
\includegraphics[width=2.5in]{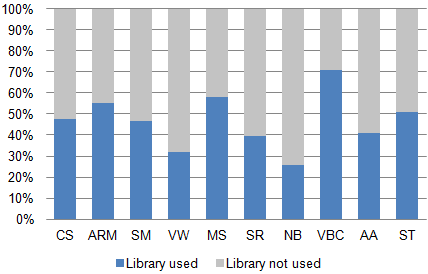}
\vspace{-.4cm}
\caption{Library usage ratio in pull requests}
\vspace{-.5cm}
\label{fig:valibpr}
\end{figure}

\textbf{Library Use in Pull Requests:} In order to provide further insight on library use, we study the pull requests of each of the projects.
Each of the requests contains one or more commits, and we analyze the changed files in those commits, and look for the occurrences of the selected libraries (Table \ref{table:library}). Fig. \ref{fig:valibpr} shows the fraction of the pull requests that use any of the 10 libraries (Table \ref{table:library}) for each project.
We note that about 50\% of the requests on average referred to those libraries in their changed files which is undoubtedly a significant amount.
For example, \emph{CS} is a large project containing 4,560 pull requests, and 47.63\% of its requests used those libraries which clearly reports an extreme dependency of the projects on the selected libraries.

Thus, to answer the first exploratory research question--Exp-RQ$_1$, commercial projects frequently use the libraries from codebase for their dedicated functionalities (\eg\ authentication, testing). 
We observe that some of the projects refer to a single library even up to 70 times, and about 50\% of their pull requests involve one or more of those libraries.

\subsection{Answering Exp-RQ$\mathbf{_2}$: Role of experience with external libraries in code review}
In order to investigate if the experience of a developer with the included libraries into a project does matter or not in the code reviews for the project,
we analyze cross-project contributions of the developers.
For each selected library from the codebase, we identify the developers who authored at least one of the merged commits, and create an \emph{author list}.
We also identify all the pull requests from each of the selected projects that use a certain library in the changed files, and then develop a \emph{reviewer list} by collecting the reviewers of the corresponding requests.
We then analyze both the author list and reviewer list, and determine if the library authors are later recommended as code reviewers or not.

\begin{figure}[!t]
\centering
\includegraphics[width=2.6in]{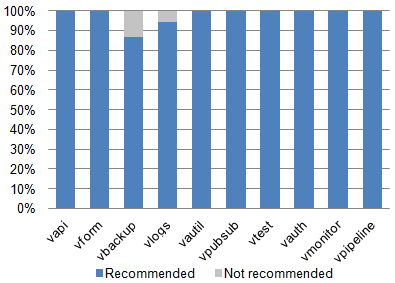}
\vspace{-.4cm}
\caption{Library authors as code reviewers for relevant pull requests in the selected projects}
\vspace{-.3cm}
\label{fig:authrev}
\end{figure}

Fig. \ref{fig:authrev} shows the percentage of the authors for each selected library (Table \ref{table:library}) who are recommended as code reviewers for different projects using that library.
We note that for each of the selected libraries except \emph{vbackup} and \emph{vlogs}, all authors (\ie\ 100\%) are later recommended as reviewers for pull requests involving those libraries. 
For \emph{vbackup} and \emph{vlogs}, such authors are also about 90\%.
Thus, the finding from our empirical dataset clearly shows that the first-hand working experience (\ie\ authorship) of a developer with relevant libraries is greatly valued in code reviewer selection, which necessarily answers our second exploratory research question-- Exp-RQ$_2$.
Furthermore, the finding also motivates \emph{cross-project experience} (\ie\ work experience with external libraries) as an expertise dimension for code review. Since the library author list largely overlaps (\ie\  98\% on average) with the reviewer list from relevant pull requests, we mostly exploit such reviewer lists in our technique for code reviewer recommendation.

%cross-project or library experience as a expertise dimension (\ie\ our proposed idea) for code reviewer recommendation.

%\begin{figure}[!t]
%\centering
%\includegraphics[width=3.3in]{techstats}
%\vspace{-.7cm}
%\caption{Relative frequency of use for specialized technologies (Table \ref{table:technology}) in VendAsta projects}
%\vspace{-.3cm}
%\label{fig:techstats}
%\end{figure}

\subsection{Answering Exp-RQ$\mathbf{_3}$: Frequency of technology use in commercial projects}
Our selected software projects are based on Google cloud platform, and they generally use \emph{Google App Engine (GAE)} technologies (Table \ref{table:technology}).
%The third research question investigates the usage frequency of different specialized technologies  in the selected projects.
In order to better understand the extent to which each of these technologies is used in those projects,
we perform our analyses using both source code files and pull requests from the projects.

\begin{figure}[!t]
\centering
\includegraphics[width=3.3in]{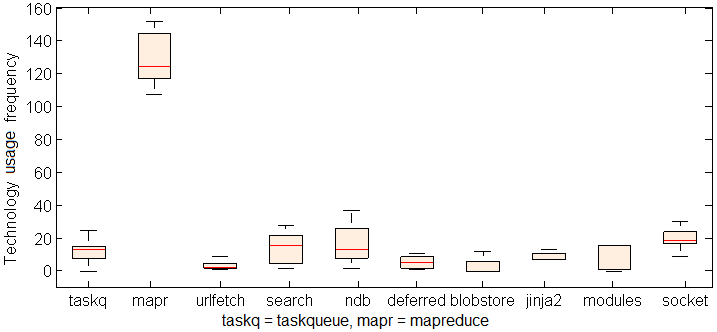}
\vspace{-.7cm}
\caption{Usage frequency of the selected technologies (Table \ref{table:technology}) in ABC Company projects (Table \ref{table:project})}
\vspace{-.2cm}
\label{fig:techfreq}
\end{figure}

\begin{figure}[!t]
\centering
\includegraphics[width=2.6in]{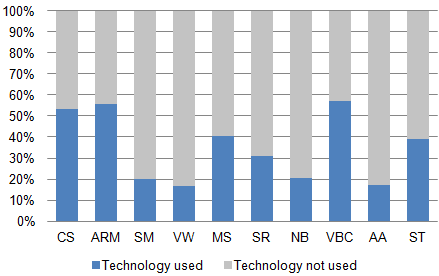}
\vspace{-.3cm}
\caption{Technology usage ratio in pull requests}
\vspace{-.6cm}
\label{fig:techpr}
\end{figure}

\textbf{Technology Use in Source Files:}
Since \emph{Google App Engine} framework targets a complete hosting solution, its service comes with a set of specialized technologies such as \texttt{taskqueue, pipeline, memcache, mapreduce} and so on. Software projects built on that framework generally use those technologies, and include them in the source files using import statements (\eg\ \texttt{from google.appengine.api import taskqueue}). 
We analyze such statements from each of the source files from the selected projects, and extract the technologies (\eg\ \texttt{taskqueue}).
 The goal is to determine the extent to which each of the selected technologies (Table \ref{table:technology}) is used in different software projects. 

Fig. \ref{fig:techfreq} shows the five point statistics on the frequency of occurrence for each of the selected technologies in different projects. 
We note that \emph{mapreduce} is the most widely used technology with a median usage frequency of about 125 and a maximum of 150.
This finding suggests that the commercial projects under study heavily use distributed computing, and \emph{mapreduce} is a major building block for them.
The other technologies such as \emph{search} and \emph{socket} also have a median frequency around 20 and a maximum frequency around 30.
Thus, the overall dependency of each of the selected projects on such technologies is noteworthy given that these technologies provide advanced or specialized computing features.
%required by Google cloud based development. 

%Fig. \ref{fig:techstats} shows the heat map for technology reference that demonstrates the comparative use of the selected technologies (Table \ref{table:technology}) in different projects.
%We note that  \emph{taskqueue, mapreduce, ndb, jinja2, modules} and \emph{socket} technologies are widely used in different projects.
%We also note that \emph{CS} project makes the maximum use of most of the technologies under study. The other projects such as \emph{MS, SR} and \emph{VBC} also adopt such technologies in a similar fashion.

\textbf{Technology Use in Pull Requests:} In order to further investigate the use of specialized technologies in the projects, we study the pull requests from each project.
We analyze the changed files from each of the pull requests, and look for the technologies referred to in the \texttt{import} statements.
Fig. \ref{fig:techpr} shows the fraction of the pull requests that refer to any of the 10 selected technologies (Table \ref{table:technology}) for each project.
We note that about 35.09\% of the requests on average used those technologies in their changed files which is undoubtedly a significant amount.
For example, \emph{CS} and \emph{VBC} are two large projects containing 4,560 and 1,050 pull requests respectively, and 53.18\% and 56.95\% of their requests used the selected technologies.
This clearly shows a significant dependency of the selected commercial software projects on the specialized technologies.

Thus, to answer the third exploratory research question-- Exp-RQ$_3$, commercial software projects frequently adopt specialized technologies such as Google App Engine technologies. We note that some of the projects refer to a single technology even up to a maximum of 150 times, and  about 35\% of their pull requests use one or more of the selected technologies. Thus, experience with such specialized technologies is also an important prerequisite for performing code review for the pull requests.

\section{C{o}RR{e}CT: Proposed Technique}\label{sec:proposed}
Since findings from the exploratory study (Section \ref{sec:explo}) suggest that experience with external software libraries or specialized technologies is
a useful proxy to code review expertise, we exploit such information in our proposed technique.  
Fig. \ref{fig:correct} shows the schematic diagram of our proposed technique-- \emph{CORRECT} for code reviewer recommendation for pull requests at GitHub. 
In our technique, we analyze the review history of past pull requests from a project, identify the relevant pull requests in terms of external library or specialized technology similarity, and then recommend code reviewers from those requests.
This section first explains our code reviewer ranking algorithm, and then discusses the implementation details of our developed prototype.

\subsection{An Overview of CORRECT}
CORRECT analyzes past pull requests and their corresponding review history from a project for ranking and then recommending code reviewers for an incoming pull request.
We believe that the developers who have reviewing experience on similar (\ie\ relevant) pull requests are suitable candidates for reviewing that request \cite{pick,reduce}.  
In this research, we hypothesize similarity between two pull requests based on their shared libraries (\eg\ \texttt{vapi, vform}) and adopted technologies (\eg\ \texttt{taskqueue, ndb}) in the changed files. Thus, any pull request, $R_i$, can be considered as a combination of the tokens for external libraries ($L_{ext}$) and specialized technologies ($T_{special}$) used in the request.
\begin{equation*}
\setlength{\abovedisplayskip}{.5em}
\setlength{\belowdisplayskip}{.5em}
R_i=\{L_i\mid L_i \epsilon L_{ext}\} \cup  \{T_i\mid T_i \epsilon T_{special}\}\\
\end{equation*}
Two pull requests ($R_i, R_{i-1}$) can be considered similar if they share a set of external software libraries or specialized technologies in the changed files.
\begin{equation*}
\setlength{\abovedisplayskip}{.5em}
\setlength{\belowdisplayskip}{.5em}
R_i \sim R_{i-1}\mid (\{L_i\}\cap \{L_{i-1}\})\not= \phi \vee (\{T_i\}\cap \{T_{i-1}\})\not= \phi
\end{equation*}

We consider 30 (\ie\ best performing heuristic count) previously closed (\ie\ merged, rejected) and similar pull requests from the code review history for analysis. It should be noted that GitHub stores the pull requests with an incremental index, and we use that index for collecting the past requests from a project. 
Our experiments also suggest that relevant requests are mostly found in a chunk of consecutive requests, and thus, we choose a list of consecutive requests from the most recent history.
We then estimate similarity degree between the current request ($R_c$) and each of the collected past requests ($R_i$) using cosine similarity measure.
We collect the library or technology names from each pull request, consider them as a \emph{bag of tokens} (\ie\ a collection of tokens with no fixed order), and decompose each token having dotted (\eg\ \texttt{app.views.filters.vff}) or underscored (\eg\ \texttt{datetime_utils}) structures. 
We then prepare a combined set of tokens, $C$, from the two sets corresponding to the two pull requests and calculate \emph{cosine similarity}, $CS(R_c,R_i)$, as follows.
\begin{equation}\label{eq:cosine}
\setlength{\abovedisplayskip}{.3em}
\setlength{\belowdisplayskip}{.5em}
CS(R_c,R_i)=\frac{\sum_{k=1}^{n}C_{ck}\times C_{ik}}{\sqrt{\sum_{k=1}^{n}C_{ck}^2}\times \sqrt{\sum_{k=1}^{n}C_{ik}^2}}
\end{equation}
Here, $C_{ck}$ represents frequency of $k^{th}$ token from $C$ in set $R_c$ (\ie\ token set from current  request), and $C_{ik}$ represents that frequency in set $R_{i}$ (\ie\ token set from the past request). 
This measure values from zero (\ie\ complete dissimilarity in libraries and technologies) to one (\ie\ complete similarity).
%helps to determine the relevance between a given pull request and a past request for potential code reviewer recommendation.
We then propagate the similarity estimates (as a proxy to review expertise) to the corresponding code reviewers ($LR$) of the past requests ($R_i$). 
\begin{equation}\label{eq:score}
\setlength{\abovedisplayskip}{.5em}
\setlength{\belowdisplayskip}{.5em}
LR[r]=\sum_{i=1}^{N} CS(R_c, R_{i})\mid reviews(r,R_{i})\wedge (c>i)
\end{equation}
Thus, according to our proposed idea, the developers who have more experience on the attached external libraries (\ie\ cross-project experience) and the adopted specialized technologies in the changed files of $R_c$, are more appropriate for code review than the ones having less experience. 

\begin{figure}[!t]
\centering
\includegraphics[width=3.3in]{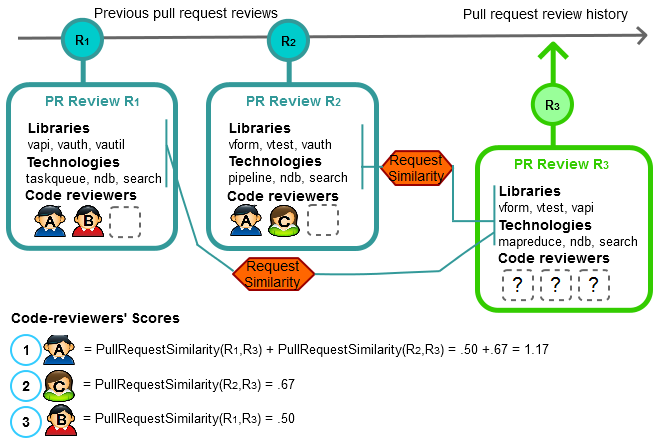}
\vspace{-.6cm}
\caption{Schematic diagram of the proposed technique-- CORRECT}
\vspace{-.5cm}
\label{fig:correct}
\end{figure}

%\subsection{Cosine Similarity}\label{sec:cosine}
% It is a measure which is frequently used in information retrieval in order to determine similarity between two text documents. 
% In our research, we use the cosine measure for determining similarity between the imported modules or libraries in the source files from two pull requests.
%The goal is to determine relevance between two pull requests based on the usage of similar external libraries and technologies, and then recommend the code reviewers of a past relevant request for the request at hand.    

%We collect the imported module names from each pull request, consider them as a \emph{bag of tokens} (\ie\ a collection of tokens with no fixed order), and decompose each token having dotted (\eg\ \texttt{app.views.filters.vff}) or underscored (\eg\ \texttt{datetime_utils}) structure. 
%We then prepare a combined set of tokens, $C$, from the two sets corresponding to two pull requests and calculate \emph{cosine similarity}, $S_{cos}$, as follows.

%Here, $A_i$ represents frequency of $i^{th}$ token from $C$ in set A (\eg\ token set from current  request), and $B_i$ represents that frequency in set B (\eg\ token set from the past request). This measure values from zero (\ie\ complete dissimilarity in libraries and technologies) to one (\ie\ complete similarity), and 
%helps to determine the relevance between a given pull request and a past request for potential code reviewer recommendation.

\textbf{Example:} 
Let us consider $R_3$ (Fig. \ref{fig:correct}) is a pull request to be submitted, and the submitter is looking for one or more
code reviewers for the request. $R_1$ and $R_2$ are two past requests similar to $R_3$ containing one or more changed files.
From Fig. \ref{fig:correct}, we note that each of $R_1$ and $R_2$ includes three libraries, adopts three specialized technologies, and is reviewed by a different set of developers. Similarly, $R_3$ also includes three libraries from the codebase and adopts three specialized technologies in the changed files.
In order to recommend reviewers for $R_3$, CORRECT first determines the \emph{cosine similarity} (Equation \ref{eq:cosine}) between libraries and technologies of $R_3$ and those of $R_1$ and $R_2$.
It then applies those scores (Equation \ref{eq:score}) to the corresponding reviewers of $R_1$ and $R_2$.
Thus, the developers who have the most review experience with similar past requests, bubble up in the ranked list for code reviewers.
From Fig. \ref{fig:correct}, we see that reviewer $A$ scores the top (\ie\ 1.17) in the list according to our ranking algorithm, and thus, A is recommended as the code reviewer for $R_3$.
We recommend the top five code reviewers \cite{reduce,pick} from such a ranked list for a pull request.
%in order to avoid old but similar requests for reviewer recommendation.   

%\setlength{\textfloatsep}{0pt}
\begin{algorithm}[!t]
\caption{Code Reviewer Ranking Algorithm}
\label{correct}
%\begin{spacing}{0.5}

\begin{algorithmic}[1]
\Procedure{CORRECT}{$R_n$}\Comment{$R_n$: new pull request}
\State $LR \gets$ \{\}\Comment{list of code reviewers}
\LineComment{VA libraries and specialized technologies used}
\State \emph{libtech} $\gets$ getLibTechTokens($R_n$)
%\Comment{collecting libraries and technologies}
%\Comment{Collecting most recent pull requests}
\LineComment{Collecting previously closed pull requests}
\State \emph{pastPRequests} $\gets$ getAllPastPRequests($R_n$)
\State \emph{pastPRequests} $\gets$ getRecentPRs(\emph{pastPRequests})
%\\\Comment{Accessing each past request}
\LineComment{Accessing \& analyzing each pull request}
\For{PullRequest $R_i\in$ \emph{pastPRequests}}
%\State libTokens$_i$$\gets$getLibTokens($R_i$)
%\State techTokens$_i$$\gets$getTechTokens($R_i$)
%\State libtechTokens$_i$$\gets$concat(libTokens$_i$,techTokens$_i$)
\State \emph{libtech}$_i$ $\gets$ getLibTechTokens($R_i$)
\\\Comment{Calculate similarity score between two requests}
\State S$_{cos}$ $\gets$ CosineSimilarity(\emph{libtech}, $libtech_i$)
\LineComment{Assigning scores to corresponding reviewers}
\State \emph{pastReviewers} $\gets$ getPRReviewers($R_i$)
\For{PR-Reviewer $r$ $\in$ \emph{pastReviewers}}
\State $LR[r]$.score $\gets$  $LR[r]$.score + $S_{cos}$
\EndFor
\EndFor
\LineComment{Creating ranked list of reviewers}
\State $RLR \gets$sortReviewersByScore($LR$) 
\State \textbf{return} $RLR$ 
\EndProcedure
\Procedure{getLibTechTokens}{$R_p$}\LineComment{$R_p$: pull request}
\LineComment{Collecting VA libraries included}
\State \emph{libs} $\gets$ getVALibTokens($R_p$)
\LineComment{Collecting specialized technologies adopted}
\State \emph{techs} $\gets$ getSpecTechTokens($R_p$)
\LineComment{Returning combined token list}
\State \textbf{return} concatTokens($libs, techs$)

\EndProcedure
\end{algorithmic}
%\end{spacing}
%\afterpage{\global\setlength{\textfloatsep}{\oldtextfloatsep}}
\end{algorithm}

\subsection{Code Reviewer Ranking Algorithm}
The pseudo-code of our proposed ranking algorithm-- CORRECT is given in Algorithm \ref{correct}. 
The algorithm takes a pull 
request--$R_n$ as an input and returns a ranked list of code reviewers--$RLR$ as the output. 
First, the algorithm extracts the \emph{external software libraries} included and the \emph{specialized technologies} used in the changed files of $R_n$ (Line 4) by invoking another procedure--\texttt{getlibTechTokens($R_{p}$)} (Line 23 to Line 30).
It then collects the most recent and previously closed pull requests from the project (Line 6 and Line 7).
%While existing studies \cite{pick} analyze all previous requests from the history, we found that computationally expensive in the case of a large history.
Our iterative experiments on ABC dataset suggest that past 30 pull requests (in contrast to all requests from the history \cite{pick}) are enough to sufficiently recommend the code reviewers. 
We thus collect the top 30 pull requests from the most recent request history. 
The algorithm then browses through each ($R_i$) of the past requests, extracts their libraries and technologies, and determines similarity ($S_{cos}$) between $R_i$ and $R_n$ using cosine similarity measure (Line 9 to Line 12). 
It then collects the corresponding reviewers of $R_i$, and assigns the score--$S_{cos}$ to each of those reviewers. 
Thus, the frequent reviewers from the similar (\ie\ relevant) pull requests get the maximum scores (Line 14 to Line 17).
Since the exploratory study (Section \ref{sec:explo}) suggests library experience and technology experience as useful proxies to code review expertise, and we identify relevant past requests using that information,
the heuristic expertise scores ($S_{cos}$) are generally propagated to the frequent and expert reviewers. 
The reviewers' names and their scores are stored in a key-value pair list--$LR$.
Once the outer loop (Line 9 to Line 18) terminates, $LR$ is sorted by score, and top five code reviewers (as suggested by literature \cite{reduce,pick} and \emph{ABC} developers) from the ranked list--$RLR$ are recommended for $R_n$ (Line 20 and Line 21).

\subsection{CORRECT's Architecture}
We adopt a client-server architecture in the implementation of our code reviewer recommendation technique. It has two parts--\emph{CORRECT server} and \emph{CORRECT client}.
The server is hosted as a web service, and it has API access to ABC code repositories at GitHub.
%responds to the requesting client, a Google Chrome plug-in.
%accepts requests for recommendation and 
%The server has access both to online (\eg\ GitHub API) and offline (\eg\ GitHub data dump) data, and it is responsible for training and updating the \emph{developer expertise model}.
The client is implemented as a Google Chrome plug-in, and it can request the server for recommendation.
Once the client plug-in captures necessary information (\eg\ details of the branch to be merged) from a pull request to be submitted, 
it encodes the information and sends to the server using an AJAX call.
The server analyzes the recommendation request, executes the proposed recommendation algorithm (Algorithm \ref{correct}), and then returns a ranked list of code reviewers to the client.
%expert and appropriate developers for code review. 
%The reviewers are then recommended by the plug-in in the context of the pull request at GitHub. 
Both client and server modules are available online for replication or third party use \cite{correct}.
%Fig. \ref{fig:corplugin} shows a simple prototype of our Chrome plug-in intended for code reviewer recommendation. Right now the recommendation is shown in a pop up window; however, we are working to show the ranked list as a context menu (\eg\ Fig. \ref{fig:motiv}) within the pull request comment box.

%\begin{figure}[!t]
%\centering
%\includegraphics[width=3.3in]{corplug3}
%\vspace{-.6cm}
%\caption{Prototype of CORRECT client, a Google Chrome plug-in}
%\vspace{-.3cm}
%\label{fig:cp}
%\end{figure}

\section{Experiment} \label{sec:experiment}
One of the most effective ways for evaluating a code reviewer recommendation technique is to consult with actual code reviews and the reviewers assigned for them from a codebase.
%We evaluate our technique using 13,081 pull requests and their code review details from VendAsta codebase. 
%We use a sliding window based selection of requests  where a sliding window approach is used on pull request history for evaluation.
%In order to validate our technique, we also compare with one state-of-the-art technique, and experiment with six open source projects from GitHub.
We evaluate our technique using the real code reviews data from ABC codebase. 
In particular, we use 13,081 pull requests and their code review details from ABC Company as our oracle in evaluating CORRECT against a number of popular performance metrics.
In order to further validate our findings and demonstrate its superiority, we experiment with six open source systems of three different programming languages, and compare with the state-of-the-art technique. 
We particularly answer the following research questions through our conducted experiments: 
\begin{itemize}[noitemsep,topsep=1pt]
%\item \textbf{RQ$\mathbf{_1}$:} Can relevance between two pull requests in terms of shared external libraries and specialized technologies be exploited in code reviewer recommendation?
\item \textbf{RQ$\mathbf{_1}$:} How does our technique--CORRECT perform in terms of the state of the art performance metrics?
%\item \textbf{RQ$\mathbf{_2}$:} Are the \emph{cross-project (\ie\ external library) experience} and the \emph{specialized technology experience} useful proxies for code review skills? 
\item \textbf{RQ$\mathbf{_2}$:} Does CORRECT outperform the state of the art technique for reviewer recommendation?
\item \textbf{RQ$\mathbf{_3}$:} Does CORRECT perform equally on both private and public codebase?
\item \textbf{RQ$\mathbf{_4}$:} Does CORRECT show bias to any of the development frameworks?
\end{itemize}

\begin{table}
\centering
\caption{Experimental Dataset (ABC Company)}\label{table:dataset}
%\vspace{-.2cm}
\resizebox{3.3in}{!}{%
\begin{threeparttable}
\begin{tabular}{l|c|c|c||l|c|c|c}
\hline
\textbf{Project} & \textbf{\#TPR}\tnote{1} & \textbf{\#SPR}\tnote{2} & \textbf{\#RPR}\tnote{3}& \textbf{Project} & \textbf{\#TPR}& \textbf{\#SPR} & \textbf{\#RPR}\\
\hline
CS & 4,560 & 3,370 & 5 & SR  & 1,927 & 1,771 &2\\
\hline
ARM  & 969 & 867 &2 & NB & 828 & 731 &2 \\
\hline
SM  & 1,291 & 1,199 &2 & VBC & 1,050 & 906 & 2 \\
\hline
VW  & 787 & 768 &1 & AA  & 1,313 &1,159 & 2 \\
\hline
MS  & 1,156 & 1,092 &2 & ST & 1,397 & 1,218 & 2\\
\hline
\end{tabular}
%\begin{tablenotes}
%\item [1] No. of example pairs for which relative quality evaluation matches with that of StackOverflow
%\item [2] \% of agreement, \item [3] \% of disagreement
 %\end{tablenotes}
\center
$^1$Total pull requests, $^2$Selected pull requests, $^3$Reviewers per request
\end{threeparttable}
%\vspace{-.2cm}
}
\vspace{-.4cm}
\end{table}

\begin{table}[!t]
\centering
\caption{Experimental Dataset (Open Source)}\label{table:ossproject}
%\vspace{-.2cm}
\resizebox{3.3in}{!}{%
\begin{threeparttable}
\begin{tabular}{l|l|c|c||l|l|c|c}
\hline
\textbf{Project} & \textbf{Lang.\tnote{1}} & \textbf{\#PR}\tnote{2} & \textbf{\#PRR}\tnote{3}& \textbf{Project} & \textbf{Lang.} & \textbf{\#PR} & \textbf{\#PRR}\\
\hline
Beets & Python & 476 & 44 & St2 & Python & 548 & 14\\
\hline
Orientdb & Java & 283 & 22 & Okhttp &Java & 650 & 49 \\
\hline
Rubocop & Ruby & 860 & 86 & Vagrant & Ruby & 1,217 & 546 \\
\hline
\end{tabular}
%\begin{tablenotes}
%\item [1] No. of example pairs for which relative quality evaluation matches with that of StackOverflow
%\item [2] \% of agreement, \item [3] \% of disagreement
 %\end{tablenotes}
\center
$^1$ Programming language  $^2$Total pull requests, $^3$Total pull request reviewers
\end{threeparttable}
%\vspace{-.2cm}
}
\vspace{-.4cm}
\end{table}

\begin{table*}
\centering
\caption{Experimental Results with Individual Subject Systems (ABC Company)}\label{table:result}
%\vspace{-.2cm}
\resizebox{6.8in}{!}{%
\begin{threeparttable}
\begin{tabular}{l|c|c|c|c|c|c|c|c|c|c|c}
\hline
 & \textbf{CS} & \textbf{ARM} & \textbf{SM} & \textbf{VW} &\textbf{MS} & \textbf{SR} & \textbf{NB} & \textbf{VBC} & \textbf{AA} & \textbf{ST} & \textbf{Avg.}\\
\hline
\textbf{Top-K Accuracy} & 93.92\% & 96.31\% & 96.75\% & 97.92\% & 94.51\% & 97.85\% & 68.81\% & 88.52\% & 89.82\% & 97.13\% & \textbf{92.15}\% \\
\hline
\textbf{Mean Reciprocal Rank (MRR)} & 0.73  & 0.66 & 0.63 & 0.89 & 0.65 & 0.71 & 0.47 & 0.57 & 0.62 & 0.74 & \textbf{0.67}\\
\hline
\textbf{Mean Precision (MP)} & 75.01\% & 91.90\% & 90.06\% & 97.79\% & 86.29\% & 93.66\% & 65.27\% & 82.81\% & 82.53\% & 93.94\% & \textbf{85.93}\%\\
\hline
\textbf{Mean Recall (MR)} & 65.56\% & 87.85\% & 85.80\% & 96.79\% & 83.25\% & 89.14\% & 60.71\% & 78.22\% & 75.94\% & 90.64\% & \textbf{81.39}\%\\
\hline
\end{tabular}
%\begin{tablenotes}
%\item [1] No. of example pairs for which relative quality evaluation matches with that of StackOverflow
%\item [2] \% of agreement, \item [3] \% of disagreement
 %\end{tablenotes}
\center
%$^1$Total pull requests, $^2$Selected pull requests, $^3$Reviewers per request
\end{threeparttable}
%\vspace{-.2cm}
}
\vspace{-.4cm}
\end{table*}

\begin{table*}
\centering
\caption{External Library Similarity \& Specialized Technology Similarity}\label{table:libtech}
%\vspace{-.2cm}
\resizebox{5.1in}{!}{%
\begin{threeparttable}
\begin{tabular}{l|c|c|c||c|c|c||c|c|c}
\hline
 & \multicolumn{3}{c||}{\textbf{Library Similarity}} & \multicolumn{3}{c||}{\textbf{Technology Similarity}} & \multicolumn{3}{c}{\textbf{Combined Similarity}}\\
\hline
& \textbf{Top-1} &\textbf{ Top-3} & \textbf{Top-5}  & \textbf{Top-1} &\textbf{ Top-3} & \textbf{Top-5}  & \textbf{Top-1} &\textbf{ Top-3} & \textbf{Top-5}  \\
\hline
\textbf{Top-K Accuracy} & 51.10\% & 83.57\% & 92.02\%  & 46.55\% & 82.18\% & 91.83\%  & 49.58\% & 83.75\% & 92.15\%  \\
\hline
\textbf{MRR} & 0.51 & 0.66 & 0.67  & 0.47 & 0.62 & 0.64  & 0.50 & 0.65 & 0.67 \\
\hline
\textbf{MP} & 51.10\% & 65.93\% & 85.28\%  & 46.55\% & 62.99\% & 83.93\% & 49.58\% & 65.98\% & 85.93\% \\
\hline
\textbf{MR} & 24.10\% & 58.34\% & 80.77\%  & 21.99\% & 55.77\% &79.50\%  & 23.29\% & 58.43\% & 81.39\% \\
\hline
\end{tabular}
%\begin{tablenotes}
%\item [1] No. of example pairs for which relative quality evaluation matches with that of StackOverflow
%\item [2] \% of agreement, \item [3] \% of disagreement
 %\end{tablenotes}
\center
%$^1$Total pull requests, $^2$Selected pull requests, $^3$Reviewers per request
\end{threeparttable}
%\vspace{-.2cm}
}
\vspace{-.4cm}
\end{table*}

\subsection{Experimental Dataset}\label{sec:dataset}
We use 10 projects (Table \ref{table:dataset}) from ABC codebase for our experiments.
It should be noted that these projects were also selected for the exploratory study (Section \ref{sec:explo}).
Since they met certain important constraints (\ie\ details in Section \ref{sec:dscoll}) for that study, they are also suitable subject systems for our evaluation.
It should also be noted that the experiments involve actual recommendation of code reviewers, and choice of the same systems does not have 
any impacts on the evaluation since there is no direct relation with the exploratory study. 
Rather, choosing the same systems confirms the findings of the exploratory study and vice versa.
The selected systems are based on Google cloud platform, and they focus on different business functions such as business reputation management, social marketing, sales and brand analytics.
Each of these projects is chosen carefully, and project related data are collected using Git Bash and GitHub API.
%It should be noted that all of the pull requests from each project were not used for experiments. We found several pull requests that neither contained any source files nor assigned any code reviewers.
%Such requests were discarded, and we chose a collection of 13,081 pull requests for the experiments.
We apply a sliding window approach \cite{sliding} (\ie\ window size=30) in selecting past pull requests from the request history to collect candidate reviewers for each current request.
In \emph{ABC Company}, one generally assigns code reviewers for a pull request by referring one or more peers in the message body of the request (\eg\ @smelnyk-va). 
We collect such developer references (\ie\ recommended reviewers), and make a \emph{reviewer set} for each pull request. 
Besides, we also identify the developers who posted feedback against the request (\ie\ actual reviewers) from the comment history of the request (Fig. \ref{fig:mcr}), and they are 
appended to the reviewer set. We then use such set as the \emph{gold reviewer set} for each of the pull requests in our experiments. 

We also select six open source projects for our experiment written in three different programming languages--\emph{Python, Java} and \emph{Ruby}.
Each project had at least 275 closed pull requests, and the corresponding gold reviewer set was developed by following the similar steps above.
Table \ref{table:ossproject} shows the summary statistics of the selected open source projects.

\subsection{Performance Metrics}\label{sec:metrics}
Since our technique focuses on recommendation, we choose two relevant performance metrics for evaluation from the corresponding literature \cite{pick,reduce,yu}. We also choose two metrics from information retrieval domain due to the inclination of this technique to this domain.

\textbf{Top-K Accuracy}: It refers to the percentage of the pull requests for which at least one reviewer is correctly recommended within the Top-K results by a recommendation technique. Top-K Accuracy can be defined as follows:
\begin{equation*}
\setlength{\abovedisplayskip}{0em}
\setlength{\belowdisplayskip}{0em}
Top{-}K Accuracy (R) = \frac{\sum_{pr\in R} isCorrect(pr, Top{-}K)}{|R|} \time 100\%
\end{equation*}
Here, $isCorrect(pr, Top{-}K)$ returns a value 1 if there exists at least one reviewer from the gold set in the $Top{-}K$ results, and returns 0 otherwise.
$R$ denotes the set of all pull requests. The higher the accuracy, the better the technique.

\textbf{Mean Reciprocal Rank (MRR)}: Reciprocal rank (RR) refers to the multiplicative inverse of the rank of the first correct result in the ranked list by a recommendation technique.
Mean Reciprocal Rank (MRR) averages such measures for all pull requests. It can be defined as follows:
\begin{equation*}
\setlength{\abovedisplayskip}{0em}
\setlength{\belowdisplayskip}{0em}
MRR(R) =\frac{1}{|R|}\sum_{pr\in R}{\frac{1}{rank(pr)}}
\end{equation*}
Here, $rank(pr)$ returns the rank of the first correct answer from a ranked list. MRR can take a maximum value of 1. The higher the MRR value, the better the technique.  

\textbf{Mean Precision (MP)}: It refers to the percentage of code reviewers which are correctly recommended for a pull request by a technique. Mean Precision (MP) averages such measures for all requests from the dataset.
%\begin{equation*}\label{eq:mp}
%\setlength{\abovedisplayskip}{0em}
%\setlength{\belowdisplayskip}{0em}
%MP= \frac{1}{|R|}\sum_{pr\in R} \frac{|common(RR_{pr},GR_{pr}, Top{-}K)|}{|RR_{pr}|}
%%P=\sum_{rq} ,~~ MP=\frac{\sum_{i=1}^{N}P_{i}}{N}
%\end{equation*}
%Here, $common(RR_{pr},GR_{pr}, Top{-}K)$ returns the common reviewers between recommended ($RR$) and gold code reviewers set ($GR$) for each pull request ($pr$). 

\textbf{Mean Recall (MR)}: It refers to the percentage of gold set reviewers which are correctly recommended for a pull request by a recommendation technique. Mean Recall (MR) averages such measures for all requests from the dataset.
%\begin{equation*}\label{eq:mr}
%\setlength{\abovedisplayskip}{0em}
%\setlength{\belowdisplayskip}{0em}
%MR= \frac{1}{|R|}\sum_{pr\in R} \frac{|common(RR_{pr},GR_{pr}, Top{-}K)|}{|GR_{pr}|}
%%P=\sum_{rq} ,~~ MP=\frac{\sum_{i=1}^{N}P_{i}}{N}
%\end{equation*}

\subsection{Evaluation with ABC Systems}
We evaluate our technique using a collection of 13,081 pull requests from 10 subject systems and four state of the art performance metrics as described in Section \ref{sec:metrics}.
We apply a sliding window based selection (window size=30) of past pull requests from the request history for each current request, and collect the code reviewer candidates.
Then the candidates are ranked by our technique based on their experience both on the external libraries and the specialized technologies used in the current request.
We then compare the ranked reviewer list with corresponding \emph{gold reviewer set} for each of the requests.
Table \ref{table:result} and Table \ref{table:libtech} summarize the performance details of our technique.
In this section, we discuss our evaluation results, and answer RQ$_1$.

Table \ref{table:result} shows the performance of our technique for each of the individual subject systems.
In this case, only top five reviewers are considered, and we note that the technique provides a recommendation accuracy around 90\% or above for all the systems except \emph{NB}.
This suggests a Top-K Accuracy of 92.15\% for our technique which is highly promising according to relevant literature \cite{pick,reduce,xin}.
Our technique also provides a Mean Reciprocal Rank of 0.67 which is quite high \cite{reduce}. 
More interestingly, on average, the recommendation technique returns results with 85.93\% precision and 81.39\% recall which suggests its greater potential for recommendation.
Thus, our technique performs significantly well in terms of all four state-of-the-art performance metrics, and the findings clearly answer our first research question, \textbf{RQ$_\mathbf{1}$}.

%It should be noted that such recommendation is based on our identification of relevant pull requests from the past review history of a project, and the relevance is determined by exploiting shared external libraries and specialized technologies between a past pull request and a request at hand.
%Thus, our findings clearly answer \textbf{RQ$\mathbf{_1}$} which suggests that 
%those similarities are effective proxies for relevance, and 
%the relevance between pull requests has a great potential for code reviewer recommendation. 

Table \ref{table:libtech} shows the performance of our technique for three different cases involving the use of similarity metrics such as \emph{external library similarity} and \emph{specialized technology similarity}. 
The goal is to demonstrate the effectiveness of those metrics as a proxy for relevance between two pull requests which provides the basis for our reviewer recommendation.  
We consider Top-1, Top-3 and Top-5 reviewers recommended  by our technique and determine the performance for each case. 
From Table \ref{table:libtech}, we note that the technique performs almost identically except for the Top-1 case when both metrics are considered in isolation.
It provides about 92\% Top-K Accuracy and above 80\% precision and recall for both metrics with Top-5 results considered.
This suggests that both similarity metrics are effective proxies for the relevance between two pull requests which drives our recommendation.
Our recommendation algorithm (Line 11 to Line 17, Algorithm \ref{correct}) exploits such similarities for estimating developer's expertise for code review which includes cross-project experience (\ie\ library experience) and specific technology experience.
%Since both similarities heuristically capture corresponding experience and are also found effective, our findings actually suggest that cross-project experience and specific technology experience are potential proxies for code review skills.
%Thus, cross-project experience and specific technology experience are potential proxies for code review skills.
Combing both metrics marginally improves the performance of our technique which also justifies the combination.
%As regard to \textbf{RQ$_\mathbf{1}$}, these findings not only suggest that our technique is highly promising in terms of various performance metrics but also explain how the technique performs so.

Our technique leverages recency of pull requests largely for code reviewer recommendation, and we also investigate its rationale using experiments. We conduct the experiments (1) using recent 30 pull requests and (2) using all available pull requests.
CORRECT provides 17\%-20\% more accuracy for the first setting with better precision, better recall and better reciprocal rank, and thus, our choice of recent pull requests for historical learning is possibly also justified.

%We also evaluated our technique involving 16 developers from \emph{VendAsta Technologies}. Each of the participants created at least two pull requests from their working projects, and our technique suggested code reviewers successfully for most of the requests.
%According to the developers, the technique is simple, easy to use and could be highly promising. One of their suggestions was real time recommendation. 
%To date, we applied multi-threading, and our technique returns results within 5--15 seconds on average for each pull request.
%While we are still improving the tool, it has been used by a number of developers from VendAsta on ad-hoc basic.
 
\begin{table}[!t]
\centering
\caption{Comparison between CORRECT and RevFinder using all Systems (ABC Company)}\label{table:comparison}
%\vspace{-.2cm}
\resizebox{3.1in}{!}{%
\begin{threeparttable}
\begin{tabular}{l|l|c|c|c}
\hline
\textbf{Technique} & \textbf{Metric} & \textbf{Top-1} & \textbf{Top-3} & \textbf{Top-5}\\
\hline
\multirow{4}{*}{RevFinder \cite{pick}} & Top-K Accuracy & 54.48\% & 76.29\% & 80.72\%  \\
\hhline{~----}
& MRR & 0.54 & 0.64 & 0.65 \\
\hhline{~----}
& MP & 54.48\%& 64.42\% & 77.24\%  \\
\hhline{~----}
& MR & 25.83\% & 57.17\% & 73.27\%  \\
\hline
\hline
\multirow{4}{*}{CORRECT} & Top-K Accuracy & 49.58\% & 83.75\% & \textbf{92.15}\% \\
\hhline{~----}
& MRR & 0.50 & 0.65 & \textbf{0.67} \\
\hhline{~----}
& MP & 49.58\% & 65.98\% & \textbf{85.93}\% \\
\hhline{~----}
& MR & 23.29\% & 58.43\% & \textbf{81.39}\%\\
\hline
\end{tabular}
\center
%$^1$Total pull requests, $^2$Selected pull requests, $^3$Reviewers per request
\end{threeparttable}
%\vspace{-.2cm}
}
\vspace{-.2cm}
\end{table}

\begin{figure}[!t]
\centering
\includegraphics[width=2.3in]{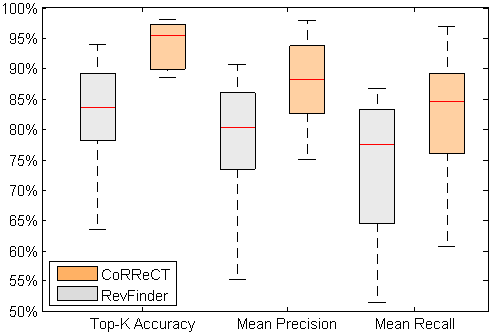}
\vspace{-.3cm}
\caption{Comparison between CORRECT and RevFinder using Box plots (ABC systems)}
\vspace{-.6cm}
\label{fig:compare}
\end{figure}

\subsection{Comparison with Existing Techniques}\label{sec:compare}
\begin{figure}[!t]
\centering
\includegraphics[width=2.8in]{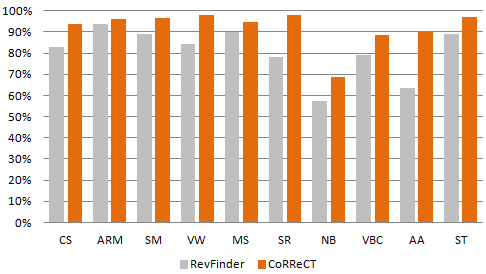}
\vspace{-.4cm}
\caption{Comparison between CORRECT and RevFinder using individual systems (ABC)}
\vspace{-.5cm}
\label{fig:comsys}
\end{figure}

In order to further validate the performance of our technique, we compare with-- RevFinder \cite{pick}, the state-of-the-art technique for code reviewer recommendation which outperformed earlier techniques as shown in their experiments.
It considers File Path Similarity \cite{rveffect} for identifying relevant reviews first and then the code reviewers.
%It outperformed the earlier techniques \cite{reduce} from the literature.
We collect authors' implementation of the competing technique, and evaluate its performance on our experimental dataset (Table \ref{table:dataset}).
Table \ref{table:comparison}, Fig. \ref{fig:compare} and Fig. \ref{fig:comsys} summarize the comparative analyses between our technique and RevFinder.

Table \ref{table:comparison} shows performance of the two recommendation techniques for Top-1, Top-3 and Top-5 results considered.
We see that our technique outperforms the competing technique in terms for all four performance metrics for Top-3 and Top-5 cases.
For example, RevFinder provides a maximum of 80.72\% recommendation accuracy with 77.24\% precision and 73.27\% recall which are comparable to the authors reported performance (\ie\ 79\% Top-K Accuracy). 
On the other hand, our technique-- CORRECT provides  a 92.15\% accuracy with 85.93\% precision and 81.39\% recall which are significantly higher.
RevFinder performs relatively better only when Top-1 result (Table \ref{table:comparison}) is considered. However, Top-1 recommendation is rarely used in practice. Rather, more than one code reviewers are generally recommended \cite{reduce,xin}.
RevFinder determines the relevance of pull requests based on merely directory structures of the changed files, which might not be always effective as suggested by our findings.
In contrast, our technique defines such relevance based on external software library or specialized technology similarity, a semantic level similarity, and the empirical findings report the potential of our technique.

Fig. \ref{fig:compare} provides further insights on the performance of the two techniques using box plots.
We collect accuracy, precision and recall of the techniques for each of the subject systems (Table \ref{table:dataset}), and derive five-point statistics for those performance metrics. 
We see that RevFinder provides a median accuracy around 85\%, a median precision of 80\% and a median recall between 75\% to 80\%.
On the other hand, our technique provides a  median accuracy over 95\%, a median precision about 90\% and a median recall about 85\%.  
Besides, the extreme limits (\ie\ maximum, minimum) are also higher for our technique than the counterparts.

Fig. \ref{fig:comsys} further validates  our technique for individual systems. We see that CORRECT outperforms RevFinder in terms of Top-K Accuracy
for each of the subject systems when Top-5 recommended reviewers are considered.
%Our technique provides the highest accuracy close to 100\% for \emph{VW} and the lowest accuracy around 70\% for \emph{NB}. Such measures for the competing technique are below 90\% and below 60\% respectively.
We perform \emph{Mann-Whitney U (MWU) test} \cite{mwu} and \emph{Cohen's d test} \cite{cohen} on the accuracy measures for checking significance and effect size respectively.
We found that the accuracy of our technique is significantly (p-value=0.003) higher than that of RevFinder. The second test returns $Cohen's~ d>1.0$ and $Glass~\bigtriangleup =0.961$ which suggest that the effect size is large, \ie\ the recommendation accuracy of our technique is largely higher than that of RevFinder.

Thus, each of the analyses above shows that our technique outperforms the state of the art--RevFinder which was found to be superior to earlier techniques \cite{pick}. This clearly answers \textbf{RQ$\mathbf{_2}$}. 
Our experimental results suggest that library or technology information is probably more effective than source file path \cite{pick} for pull request relevance which probably led to our better performance.

One might wonder why we did not compare with another relevant technique-- ReviewBot \cite{reduce} from the literature that exploits line change history of source code for reviewer recommendation.
We made that choice due to two appealing reasons. 
First, RevFinder outperforms ReviewBot by a large margin. Second, 70\%--90\% of the source code lines in the project are generally changed only once \cite{pick}. 
Thus, there is an inherent lack of sufficient line-level history, and therefore, the performance of ReviewBot is limited.

%Another recent work--\citet{ghadeer} applies code clone detection (\ie\ code similarity) in expert recommendation for programming problem solving, and is related to our work technically.
%However, our investigation reports that on average only 6.60\% of the source files from each of the subject systems contains code clone segments.
%This suggests that the pull requests also contain very little similar code segments, and thus, source code level similarity between a new request and the past requests might not be effective enough to identify the appropriate code reviewers for the request.

\subsection{Experiments with Open Source Projects}
Our recommendation technique-- CORRECT is found highly promising with an organizational codebase which is closed source and based on \emph{Python}.
In order to further validate and generalize our findings, we conduct an experiment with six open source projects from GitHub written in three different programming languages.
%--\emph{Python, Java} and \emph{Ruby}.
%We select such projects from GitHub that have at least 275 closed pull requests.
%Table \ref{table:ossproject} shows the summary statistics of the selected projects.
%In particular, we investigate the potential of our technique for reviewer recommendation on various development environments and languages, and then answer RQ$_3$ and RQ$_4$. 

\begin{table}[!t]
\centering
\caption{Comparison using Open Source Systems}\label{table:composs}
%\vspace{-.2cm}
\resizebox{3in}{!}{%
\begin{threeparttable}
\begin{tabular}{l|l|c|c|c}
\hline
\textbf{Technique} & \textbf{Metric} & \textbf{Top-1} & \textbf{Top-3} & \textbf{Top-5}\\
\hline
\multirow{4}{*}{RevFinder \cite{pick}} & Top-K Accuracy & 48.89\% & 60.87\% & 62.90\%  \\
\hhline{~----}
& MRR & 0.49 & 0.54 & 0.55 \\
\hhline{~----}
& MP & 48.89\%& 59.67\% & 62.57\%  \\
\hhline{~----}
& MR & 41.80\% & 55.86\% & 58.63\%  \\

\hline
\hline
\multirow{4}{*}{CORRECT} & Top-K Accuracy & 57.72\% & 81.01\% & \textbf{85.20}\% \\
\hhline{~----}
& MRR & 0.58 & 0.68 & \textbf{0.69} \\
\hhline{~----}
& MP & 57.72\% & 79.20\% & \textbf{84.76}\% \\
\hhline{~----}
& MR & 48.87\% & 73.44\% & \textbf{78.73}\%\\
\hline
\end{tabular}
\center
%$^1$Total pull requests, $^2$Selected pull requests, $^3$Reviewers per request
\end{threeparttable}
%\vspace{-.2cm}
}
\vspace{-.2cm}
\end{table}

\begin{figure}[!t]
\centering
\includegraphics[width=2.3in]{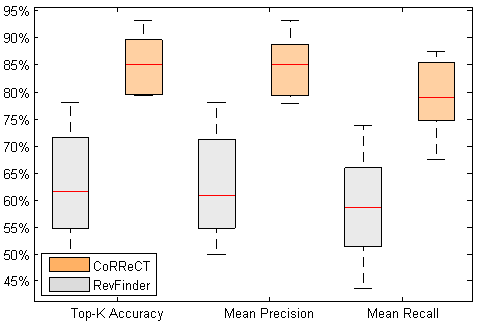}
\vspace{-.3cm}
\caption{Comparison using open source systems}
\vspace{-.3cm}
\label{fig:compare-stat}
\end{figure}

Table \ref{table:composs} shows performance details of our technique with the open source projects, and compares with the state of the art technique-- RevFinder.  
We see that RevFinder provides a maximum recommendation accuracy of 62.90\% with 62.57\% precision and 58.63\% recall
whereas our technique--CORRECT has 85.20\% accuracy with 84.76\% precision and 78.73\% recall. 
%On the other hand,our technique--CORRECT has a 85.20\% Top-K accuracy, 84.76\% precision and 78.73\% recall. 
The remaining metric--MRR is also higher for our technique. From the box plot in Fig. \ref{fig:compare-stat}, we also see that the accuracy, precision and recall of CORRECT are significantly higher.
Thus, our technique outperforms the state-of-the-art technique for open source projects as well. 

\begin{table}[!t]
\centering
\caption{Performance of CORRECT on Different Programming Languages (Open Source Systems)}\label{table:complang}
%\vspace{-.2cm}
\resizebox{3.3in}{!}{%
\begin{threeparttable}
\begin{tabular}{l|c|c||c|c||c|c}
\hline
 & \multicolumn{2}{c||}{\textbf{Python}} & \multicolumn{2}{c||}{\textbf{Java}} & \multicolumn{2}{c}{\textbf{Ruby}}\\
\hline
 & \textbf{Beets} & \textbf{St2} & \textbf{Okhttp} & \textbf{Orientdb} & \textbf{Rubocop} & \textbf{Vagrant}\\
\hline
\textbf{TKA}\tnote{1} & 93.06\%  & 79.20\% & 88.77\% & 81.27\% & 89.53\% & 79.38\%\\
\hline
\textbf{MRR} & 0.82 & 0.49 & 0.61 & 0.76 & 0.76 & 0.71 \\
\hline
\textbf{MP} & 93.06\% & 77.85\% &  88.69\% & 81.27\% & 88.49\%  & 79.17\%\\
\hline
\textbf{MR} & 87.36\% & 74.54\% & 85.33\% & 76.27\%  & 81.49\% & 67.36\%\\
\hline
\end{tabular}
\center
$^1$Top-K Accuracy
\end{threeparttable}
%\vspace{-.2cm}
}
\vspace{-.6cm}
\end{table}

Table \ref{table:complang} shows performance of our technique for each of the individual open source projects. 
We compare the performance for open source projects with that of closed source projects (Table \ref{table:result}) using MWU test and Cohen's d test.
%We found that the accuracy difference for these two project types is statistically significant. 
Although Top-K accuracy is slightly higher for commercial projects, the remaining metrics-- precision, recall and reciprocal rank are comparable.
For example, in the cases of  precision and recall, we got p-value=0.239, $Cohen's~ d=0.142$, $Glass~\bigtriangleup =0.190$ and p-value=.209, $Cohen's~ d=.276$, $Glass~\bigtriangleup =0.357$ respectively which suggest that 
such performance measures by our technique for both project types-- \emph{open source} and \emph{closed source} are not statistically different.
Thus, the findings answer \textbf{RQ$\mathbf{_3}$}, \ie\ CORRECT performs almost equally for both private and public codebases.
% with more than 85\% accuracy and precision, and about 80\% recall.

Table \ref{table:complang} also demonstrates how our reviewer recommendation technique performs with projects using various programming languages-- \emph{Python, Java} and \emph{Ruby}.
%In case of Java and Ruby, we analyze \emph{import} statements and \emph{require} statement respectively in the program files for estimating the relevance between pull requests which in turn helps estimate the experience of the code reviewers.
We did not notice any particular bias to any of the languages, and our technique provides nearly 90\% accuracy with at least one project from each of the languages.
The other performance metrics are also promising and competitive with those with \emph{ABC} projects. Thus, the findings also answer \textbf{RQ$\mathbf{_4}$} which suggests that our recommendation technique is not biased to any particular programming language (\eg\ \emph{Python}).
%for high performance. 

\section{Threats to Validity} \label{sec:threats}
Threats to internal validity relate to experimental errors and biases \cite{david}. 
Many of our subject systems (except CS) are medium sized, and they contain about 1.1K pull requests on average. 
%However,%Our experiments involve 16  subject systems, and 
%Thus, the performance details reported are mostly for medium sized systems. 
%Since we stick with an organizational codebase, and it does not contain many large projects except one (\ie\ CS), we could not evaluate our technique using large systems.
However, we consider not only more (\ie\ 10 commercial and 6 open source) systems than any of the existing studies \cite{pick,reduce} but also conduct experiments with 17K pull requests.
Besides, we performed \emph{stress testing} using 3.3K pull requests from a large open source system--\emph{Mozilla Zamboni} to deliberately make our technique failed.
However, our technique passed the test, and did not crash. 
%Manual inspection on the pull requests suggests that neither the requests are rigorously reviewed nor they contain many commits, which possibly explains the moderate performance of our technique. 
Thus, the technique is robust enough, and the dataset might be also sufficient enough to generalize our findings.

Threats to external validity relate to the generalizablity of a technique. We first experimented using only Python projects from a private codebase which might raise the generalizability concern.
In order to handle the threat, we adapt our technique for two more platforms-- Java and Ruby. From our findings with 16 (12 python, 2 Java and 2 Ruby) subject systems from three different languages, we didn't notice any bias of our technique to any project type (\ie\ open source, closed source) or any language (\ie\ Python, Java, Ruby).

Threats to construct validity relate to suitability of evaluation metrics \cite{david}. We use Top-K Accuracy and Reciprocal Rank which are widely used by relevant literature \cite{pick,reduce,yu}. 
The remaining two metrics are well known in information retrieval, and our technique is also aligned with this domain.
This  confirms no or little threats to construct validity.
%Thus, we believe that there exists little threat to construct validity.

%One may argue about the usability of the technique.
%Given that our technique is a Google Chrome plug-in that simply augments code review interface of GitHub through automated reviewer recommendation, a user study is not warranted. 
%Nevertheless, the technique was evaluated by 16 professional developers of \emph{ABC Company} for 30 minutes each. 
%The developers reported that the technique was simple, easy to use and could be highly promising.
%To date, our technique returns results within 5--15 seconds on average for each pull request which seemed okay with the developers.
%They were also satisfied with the accuracy of the technique, as was also validated with real ABC data. 
%A couple of developers suggested further work on the handling of extreme concurrent requests. 
%While we are working towards building a fully fledged tool, the current version is being used by a number of developers at ABC on an ad-hoc basis.

\section{Related Work} \label{sec:related}
\textbf{Code Reviewer Recommendation}: Existing studies recommend code reviewers by analyzing code review history-- line change history \cite{reduce} and past review comments \cite{yu, xin}, project directory structure \cite{pick, rveffect}, and developer collaboration network \cite{yu}.  
%employ different techniques such as version control history \cite{reduce, kevic, bosu}, directory structure \cite{rveffect} and authors' meta data \cite{authorship, discovery} analysis, code dependency \cite{brazil} and code similarity \cite{ghadeer} analysis  and developer communication history \cite{brazil, seafood, sna, ghadeer} analysis. 
%In summary, these approaches focus on the analysis of source file level artifacts or interactions among the developers for determining the expertise of a developer.
\citet{reduce} propose a recommendation technique--ReviewBot that analyzes change history of the affected lines in a review request.
However, existing findings show that most of the lines are generally changed only once \cite{pick} which makes the line change history really scarce and thus, the performance of  ReviewBot is limited.
\citeauthor{pick} propose another technique-- RevFinder \cite{pick} that identifies relevant review requests using File Path Similarity (FPS) \cite{rveffect}, and then recommends reviewers from those requests for a review request at hand.
RevFinder also outperformed existing techniques including ReviewBot \cite{pick}.
On the other hand, CORRECT identifies relevant pull requests using \emph{external library similarity} and \emph{specialized technology similarity} which are found to be more effective than File Path Similarity\cite{pick} for estimating relevance between pull requests, and thus for reviewer recommendation. 
In our comparative studies (Section \ref{sec:compare}), we show that our technique outperformed RevFinder with statistically significant performance improvements.
%We compare with \citet{pick}, a state-of-the-art technique, and experimental results (Section \ref{sec:compare}) show that our technique outperforms it with statistically significant performance improvement.
Another recent work \cite{xin} applied machine learning on past review comments and File Path Similarity \cite{pick}. It thus suffers from similar issues as of RevFinder such as pull request relevance issue, and that the learned models could be biased to the subject systems under study.

%The recent work of \citet{xin} improved upon \citet{pick} by applying machine learning on past review comments and File Path Similarity. Although, we did not directly compare with this technique, it suffers from the same issues of \citet{pick}, and the learned models could be biased to the subject systems under study.

The remaining technique--\citet{yu} analyzes past review comments and developer collaboration network for reviewer recommendation.
While we use library and technology similarity between pull requests for determining relevant past requests, they use review comment similarity (\ie\ textual similarity) for the same purpose.
Besides, their idea is still not properly evaluated or validated.
%\citet{rveffect} improve this algorithm where they assign reviewer points using \emph{File Path Similarity (FPS)} algorithm.
%According to \citet{rveffect}, developers who reviewed the source files from a project directory of interest are suitable for later code reviews involving that directory.
%Thus both approaches target a subset of past reviews for a later recommendation whereas our approach analyzes three types of contributions-- \emph{authorship, reviewership} and \emph{technology adoption} not only for past review selection but also for code reviewer recommendation. 

\textbf{Expert Recommendation:} \citet{ghadeer} propose an expert recommendation system that exploits \emph{code similarity} for estimating expertise of a developer on a code fragment of interest. 
%The baseline idea is, if a developer has worked on code fragments that are similar to the fragment of interest, that developer is a suitable candidate for technically assisting (\eg\ code review) with the fragment. They also apply a list of social heuristics which determine the developer's acceptance in the community, in the recommendation of expert developers. 
Similar technique is applied by \citet{brazil} where they develop a communication network among documents, source code and developers, and recommend dominant developers as experts. 
%by analyzing \emph{document-source code, developer-source code, source code-source code} and \emph{developer-developer} relationships. %They then locate the dominant developer nodes in the network as the expert developers for recommendation. 
\citet{sna} studies the developer network using code review relationship, and identify core and peripheral developers using different network properties.
%of three open source projects based on code review relationship, and analyze the contribution patterns of core developers and peripheral developers of a project. Thus while these approaches leverage various communication networks for expert developer identification, we adopt a classifier model-based approach for the same task. Furthermore, we consider different aspects of programming contributions in determining expertise for code review.    
There exist several studies in the domain of \emph{bug triaging} that analyze duplicate bug reports \cite{kevic} or apply IR-based traceability \cite{authorship} techniques for recommending experts for bug fixation.
%developers for bug fixation. \citet{kevic} identify duplicate bug reports of a reported bug by employing text similarity, and analyze the changeset of their fixations for appropriate developer recommendation. Similar approach is adopted by \citet{authorship} who use an IR-based traceability technique to locate relevant source files for a reported bug, and then exploit the authorship meta data for expert developer recommendation.
Several studies are also conducted on expert user recommendation at Stack Overflow that analyze cross-domain contributions \cite{discovery} or question difficulty \cite{difficulty} for expertise estimation.
While these expert recommendation techniques are somewhat similar to ours, their context of recommendation is different and thus, comparing ours with them is not feasible. 
Of course, we introduced two novel and effective expertise paradigms (\emph{cross-project experience} and \emph{specialized technology experience}) which were not exploited by any of the recommendation systems. This makes our technique significantly different from all of them.

 %Thus, the above techniques recommend experts in different contexts than ours, and a direct comparison is not required.
%for programming problem solving, bug fixation or question answering,
%In contrast, our technique performs quite a similar job in the context of code review by applying two novel and effective expertise paradigms.
%recommends experts for code review at GitHub.

%On the other hand, we analyze three types of programming contributions-- \emph{authorship, reviewership} and \emph{technology adoption} of a developer against the code changes in a pull request, and determine her expertise in code review.

%Thus, while the above studies focus on recommending experts for programming problem solving, our technique focus on a more specific task such as code review.
%\balance
\balance

\section{Conclusion \& Future Work} \label{sec:conclusion}
To summarize, we propose a novel technique-- CORRECT for code reviewer recommendation for pull requests at GitHub. It heuristically captures the experience of a developer with the external libraries (\ie\ cross-project experience) and specialized technologies used in a pull request for reviewer recommendation.
Experiments using 13,081 pull requests from 10 subject systems from an organizational code repository show that our technique recommends code reviewers with 92.15\% Top-5 accuracy, 85.93\% precision and 81.39\% recall which are highly promising. 
Experiments using 4,034 pull requests from six open source projects suggest that our technique performs well both for open source and closed source projects, and is not biased towards any programming languages.
%A user study with professional developers confirmed our empirical findings and the good usability of the technique.
%We package our recommendation technique into a web service and a Chrome plugin prototype. 
Comparison with one of the state of the art techniques also demonstrates the superiority of our technique.
In future, we plan to work on handling more concurrent recommendation requests as suggested by ABC developers.

%to adapt our pull request relevance idea that exploits library or technology similarity, for other types of tasks such as bug triaging and concept location.   

%To summarize, in this paper, we propose a code reviewer recommendation technique that estimates expertise of a developer by analyzing not only her activities on the \emph{changeset} associated with an incoming pull request but also her 
%domain-specific and cross-project contributions within the code base. We evaluate our proposed expertise model using a case study with XXX pull requests from the codebase of VendAsta Technologies,
%and implement the recommendation system as a Google Chrome plug-in.
%We apply our expertise model in an organizational context-- \emph{VendAsta Technologies}, and evaluate the model using a case study with XXX pull requests.

%\newpage
%adding the bibliography
\bibliographystyle{plainnat}
\setlength{\bibsep}{0pt plus 0.2ex}
\scriptsize
\bibliography{sigproc}
\end{document}